\documentstyle[twoside,fleqn]{an-art}
\input psfig.sty
\begin{document}
\input{psfig.sty}
\def\address #1END {{\vspace{9mm}\noindent\small Addresses of the authors: 
\medskip \\ #1}}
\def\addresses #1END {{\vspace{9mm}\noindent\small Addresses of the authors: 
\medskip \\ #1}}
\begin{titlepage}
\setcounter{page}{1}
\def\makeheadline{\vbox to 0pt{\vskip -30pt\hbox to 50mm
{\small Astron. Nachr., IN PRESS  \hfill}}}
\makeheadline
\title {Early versus late type galaxies in compact groups}
\author{{\sc A.A. Shaker}, Helwan-Cairo,Egypt \\
\medskip
{\small National Research Institute of Astronomy and Geophysics
(NRIAG) } \\
\noindent
{\small Osservatorio Astronomico di Capodimonte, Napoli}\\
\bigskip
{\sc G. Longo, P. Merluzzi}, Napoli, Italy\\
\medskip
{\small Osservatorio Astronomico di Capodimonte}
}
\date{Received ~~~~~~~~~~~; accepted ~~~~~~~~~~~} 
\maketitle
%
\summary
END

\end{titlepage}

\noindent
We find a strong correlation between the effective radius of the largest
early-type galaxies in compact groups of galaxies and the velocity
dispersion of the groups. The lack of a similar correlation for late type
galaxies is supportive of the so called second generation merging scenario
which predicts that ellipticals should dominate the internal dynamics of the
groups, while late-type galaxies are mainly recent interlopers which are
still in an early stage of interaction with the group potential.\\

\keyw
galaxies - galaxy groups - Hickson groups
END
\AAAcla
160
END

\vskip 0.3in
\noindent
{\bf 1.\hspace{2cm} Introduction}

\bigskip
\noindent
Compact groups of galaxies (hereafter CGs) have low velocity
dispersions (compared to clusters), small size and high spatial density and
therefore short crossing and coalescence lifetimes (in the average, $\tau
_{cr}\sim 0.016\;T_o$ where $\tau _{cr}$ and $T_o\;$are the crossing and the
Hubble time, respectively; Hickson et al. 1992). In the simple assumption
that luminous matter is a good tracer of the group potential, these
properties call for a high merging rate and turn CGs into putative factories
of elliptical galaxies. Short coalescence lifetimes, however, hit against
several observational evidences:\ the relatively large number of CGs
observed in the nearby universe leads to an expected number of ongoing or
recent mergers which is one order of magnitude higher than what it is
actually observed;\ ellipticals in CGs are on the average very luminous
objects ($28\%$ of them is more luminous than M87, Mendes de Oliveira and
Hickson 1991) but, in a given group - the sum of the luminosities of the
members is too high compared to that of luminous ellipticals;\ most
ellipticals in CGs belong to the so called ''bright'' family (Capaccioli et
al. 1992) which is formed by highly evolved objects which have experienced
at least one major (or several minor) merging events (Caon et al. 1994),\ a
fact which seems to be contradicted by the small fraction of E-type galaxies
($7\%\;$of the total sample)\ having the abnormally blue colors which are
predicted for recent (age $<5\cdot 10^8$ years, Zepf et al. 1991) mergers.
It needs to be stressed, however, that using the same data but different
assumptions, Zepf (1993) and Hickson (1997a) derive completely different
estimates for the merging rate ($7\%$ and $23\%$, respectively). The latter
value is in better agreement with the fact that in several nearby groups
faint luminous envelopes encompassing the whole group and thus suggestive of
an ongoing merger have been recently detected (Ribeiro et al. 1996, Paramo
and V\'{i}lchez 1997, Longo et al. in preparation).\\

\smallskip 
These seemingly conflicting results led many authors to suggest
that CGs could actually be physically unbound systems:\ either chance
alignments of galaxies in the field (Mamon 1987, 1990) or in filaments
(Hernquist et al. 1995, Ostriker et al. 1995), or transient configurations
within richer and more dispersed groups of galaxies (Diaferio et al. 1994).
All these models, however fail to explain one or more of the observed
properties of CGs (cf. Hickson 1997b)\ such as:\ the higher than average
frequency of interactions (estimates range from $43\%$ Mendes de Oliveira
and Hickson 1994, to $\sim 60\%$ Rubin et al. 1991); the different
morphological composition of CGs with respect to the field and to clusters;\
the recent X rays surveys which detect diffuse X-ray emission in most groups
(e.g. Pildis et al. 1995, Saracco and Ciliegi 1995, Ebeling et al. 1994)
with total X-ray luminosity correlating with both the hot gas temperature
and the velocity dispersion of the groups (Ponman et al. 1996). 
A possible way out from this sort of ''stall'' is the so called
''second generation merger scenario'' (hereafter SGS) proposed by Governato
et al. (1996). Since most groups are found to be not isolated but contained
inside larger structures such as loose (or poor) groups and clusters (Vennik
et al. 1993), the CGs we see today may be just aggregating from the
environment and beginning to interact. The first generation of these groups
was destroyed by rapid merging, stripping of dark-matter haloes and
subsequent production of a number of early merger remnants which would
appear today as early-type galaxies (overluminous and belonging to the
''bright'' family as it is actually observed) either isolated or in groups
(which would explain the concordance of morphological types observed in most
groups). These remnants would be surrounded by diffuse haloes both luminous
(stripped stars) and dark, and new arrivals would soon begin to interact
(high frequency of interaction) but would mostly avoid merging since the
collisional cross-sections for individual galaxies are now smaller
(Governato et al. 1996). SGS finds also support in the properties of the
diffuse X-ray haloes where\ i)\ the position angle of the photometric major
axis of the central early-type object is aligned to that of the diffuse X
halo ( Mulchaey and Zabludoff 1997); ii)\ the position of the dominant
early-type galaxies usually coincides with that of the photometric
baricenter of the group (as defined in Zabludoff and Mulchaey 1997). In
this paper we discuss some additional optical evidences that bright
ellipticals and lenticulars largely determine the shape of the galaxy
potential as it is predicted by SGS.\\

\smallskip
The catalogue by Hickson (1982, hereafter HCGs) is so far the most complete,
homogeneous and best studied sample of compact groups of galaxies. One of us
(AAS) as a part of his Ph.D., has compiled all published photometric (at all
wavelengths), morphological and geometrical information on HCG's in the
Hickson's list, homogenized and reduced (whereas needed) to a standard
system. This database will be used in the following discussion and we
provide here some information only on the data which are needed for the
present work. In what follows, we study the relations between the linear
effective radius $R_e$ (for early type objects only), the isophotal diameter
for all the objects $D$ and the velocity dispersion $\sigma $ of the groups
in the 92 HCGs which have at least three accordant redshift members (Hickson
et al. 1992). We rejected from our sample Hickson 54 which is more likely a
galaxy shred in pieces rather than a true group (Williams and van Gorkom
1988). Effective radii and morphological types for the early type
objects are taken from Bettoni $\&$ Fasano (1993), Fasano $\&$ Bettoni
(1994), Zepf $\&$ Whitmore (1993), and were homogenized as in Caon et al.
(1994). Among the 91 groups there are 42 for which effective radii for the
largest early-type (not necessarily dominant) galaxies (=LEG) are available
(H45d was also rejected due to its discordant redshift and H90c is not LEG\
in its group, Longo et al. in preparation). Isophotal diameters (at $\mu _B$
= 24.5 mag arcsec$^{-2}$) and asymptotic magnitudes corrected for internal
and external extinction for all the galaxies in the 91 HCGs were taken from
Hickson et al. (1989). Radial velocity dispersions and average redshifts for
the groups are from Hickson (1994). Both effective radii and isophotal
diameters were converted in linear sizes by assuming $H_o=100\,km\,s^{-1}%
\;Mpc^{-1}$. Our final sample consisted of 91 HCGs (complete sample), 42 of
which hosting at least one elliptical galaxy (LEG\ sample).

\bigskip
\noindent
{\bf 2.\hspace{2cm}Discussion}\\

\noindent
As already mentioned before, one obvious implication of the SGS is the
existence of a dichotomy of properties between early and late type objects
in the groups:\ most Es in CGs must be highly evolved objects (i.e. relics
from first generation mergers) shaping the potential well of the groups,
while late type members are just now infalling, are not yet virialized and
do not affect much the dynamics of the groups. By comparing the properties
of galaxies in the complete sample and in the LEG sample we do not find any
systematic differences between the average velocity dispersions for the
early and late type objects. The absence of a systematic difference between
the two groups, however, is likely due to the fact that groups have
different total luminosities (hence masses) and the distances from the
centers of the groups of individual galaxies should be rescaled. This
normalization, however is prevented by the small number statistics.\\

A possible evidence in favor of SGS comes instead from the linear sizes of
the brightest galaxies.\\

In Fig.1 we plot the Log of the effective radii of
the early-type objects in our LEG sample versus the velocity dispersion of
the group:\ ellipticals are shown as filled symbols and lenticulars as
asterisks;\ symbols encircled denote that the object belongs to an E-type
group (i.e. early type dominated group, Hickson 1982), while symbols
surrounded by a square mark objects belonging to S-type (late type
dominated)\ groups. The existence of a strong correlation (confidence level
larger than $99.5$ $\%$) between the two quantities is apparent and\ a
simple bisector fit gives the following equation:

\bigskip

\begin{equation}
Log\;\sigma =0.97(\pm .12)\;Log\;R_e+\;1.65(\pm 0.10)  \label{Eq. 1}
\end{equation}


\begin{figure}
\begin{center}
\hspace{0.3cm}\psfig{figure=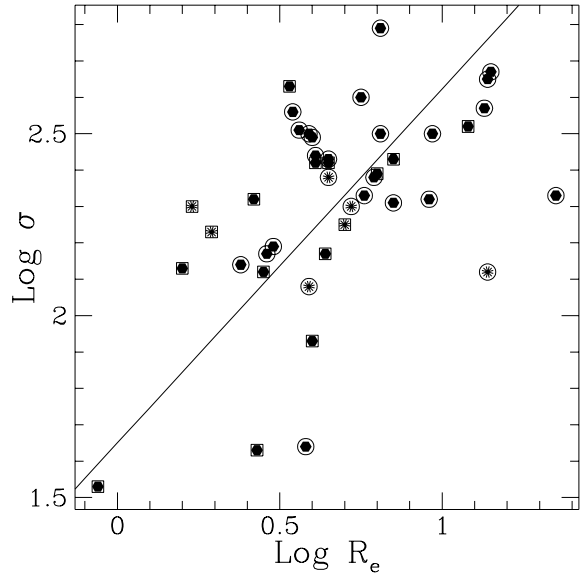,height=8cm,width=8cm}
\end{center}
\caption{Plot of the log of
the linear effective radius for the galaxies in the LEG sample versus the
velocity dispersion of the group. Symbols are as follows: filled =
ellipticals, asterisks = lenticulars. Encircled and ''in squared'' symbols
denote galaxies which belong to E-type and S-type groups, respectively. The
solid line gives the best fit.}
\end{figure}

\bigskip

The existence of a correlation between an individual ($R_e$) and a global ($%
\sigma $) property confirms that at least the groups in the LEG sample are
physically bound systems. When coupled to the well known correlation between 
$Log\;R_e$ and the total luminosity (hence the mass) for early type galaxies
(Hamabe and Kormendy 1987, Capaccioli et al. 1992) Eq. 1 implies that the
gravity potential of the groups in the LEG\ sample is dominated by the early
type objects. For what late-type objects were concerned, since effective
radii are poorly defined, we preferred to use isophotal diameters which are
expected to be shortened by the effects of tidal stripping which truncates
the disks of late type galaxies which move into high spatial density
environments (e.g. Boselli et al. 1996).

Previous attempts to detect such an effect in CGs were made by Hickson et
al. (1977), who using a non homogeneous sample of 18 groups measured on
POSS-I plates, found a correlation between the size of the largest galaxies
and the mean separation of galaxies within the group;\ by Whitmore (1990,
1992) who found a weak correlation between the size of the largest galaxy
and the spatial number density of the group, and by Maia and da Costa (1990)
who also found a correlation between size and spatial density but explained
it as partially due to distance dependent selection effects. In our sample
we do not find any correlation between $\sigma \;$and the logarithm of the
linear isophotal diameter $D$ for late type galaxies in the complete sample
(see Fig.2).

\begin{figure}
\vfill \begin{minipage}{.47\linewidth}
\begin{center}
\mbox{\psfig{figure=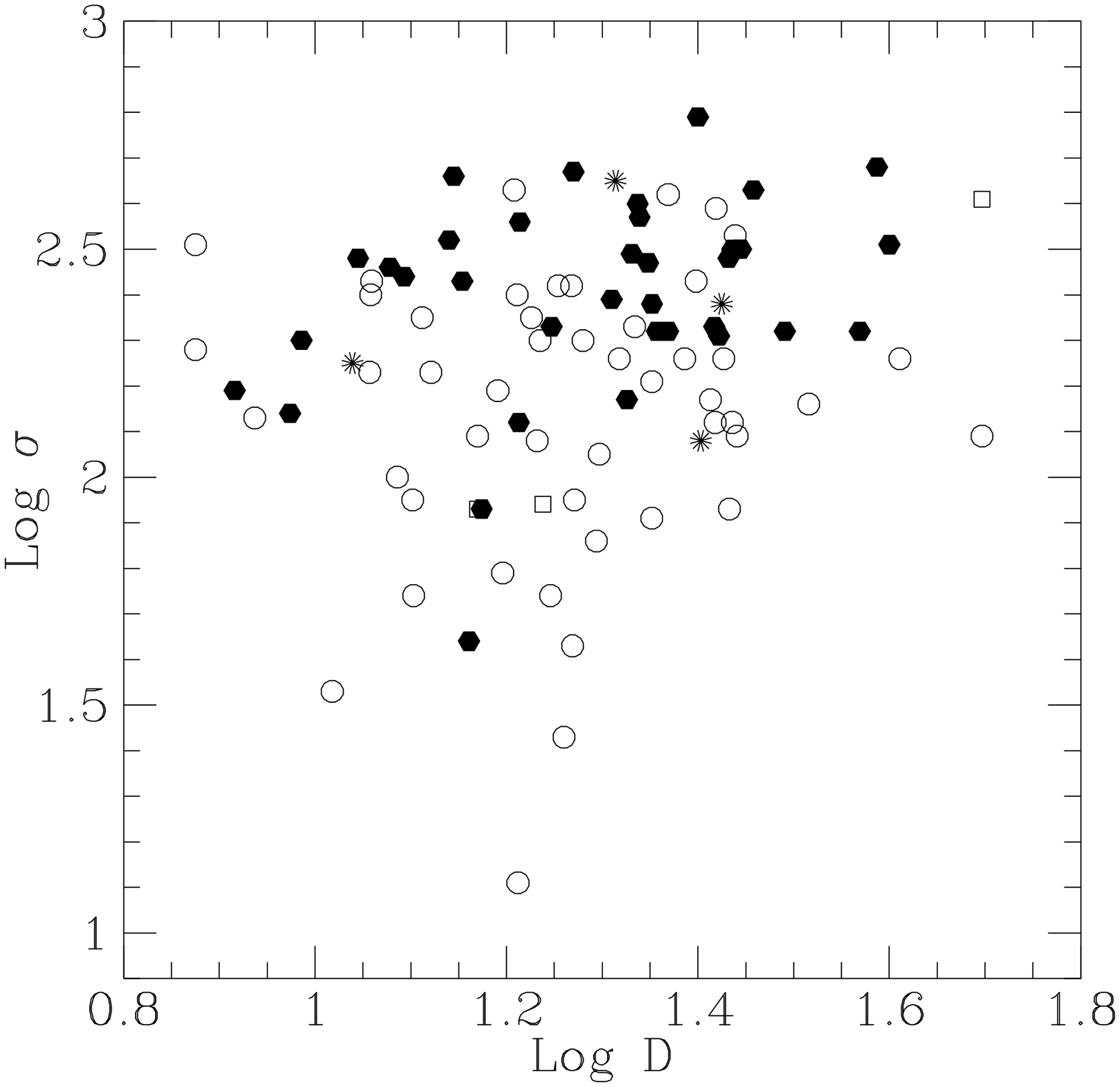,height=8cm,width=8cm}}
\caption{$Log$ $D$ vs $Log$ $\sigma $
for galaxies in the complete sample. Symbols are as follows: filled =
ellipticals, asterisks = lenticulars, empty circles = spirals, empty squares
=\ irregulars.}
\end{center}
\end{minipage} 
\hspace{-0.7cm} \hfill 
\begin{minipage}{.47\linewidth}
\begin{center} 
\vspace{-0.4cm}\mbox{\psfig{figure=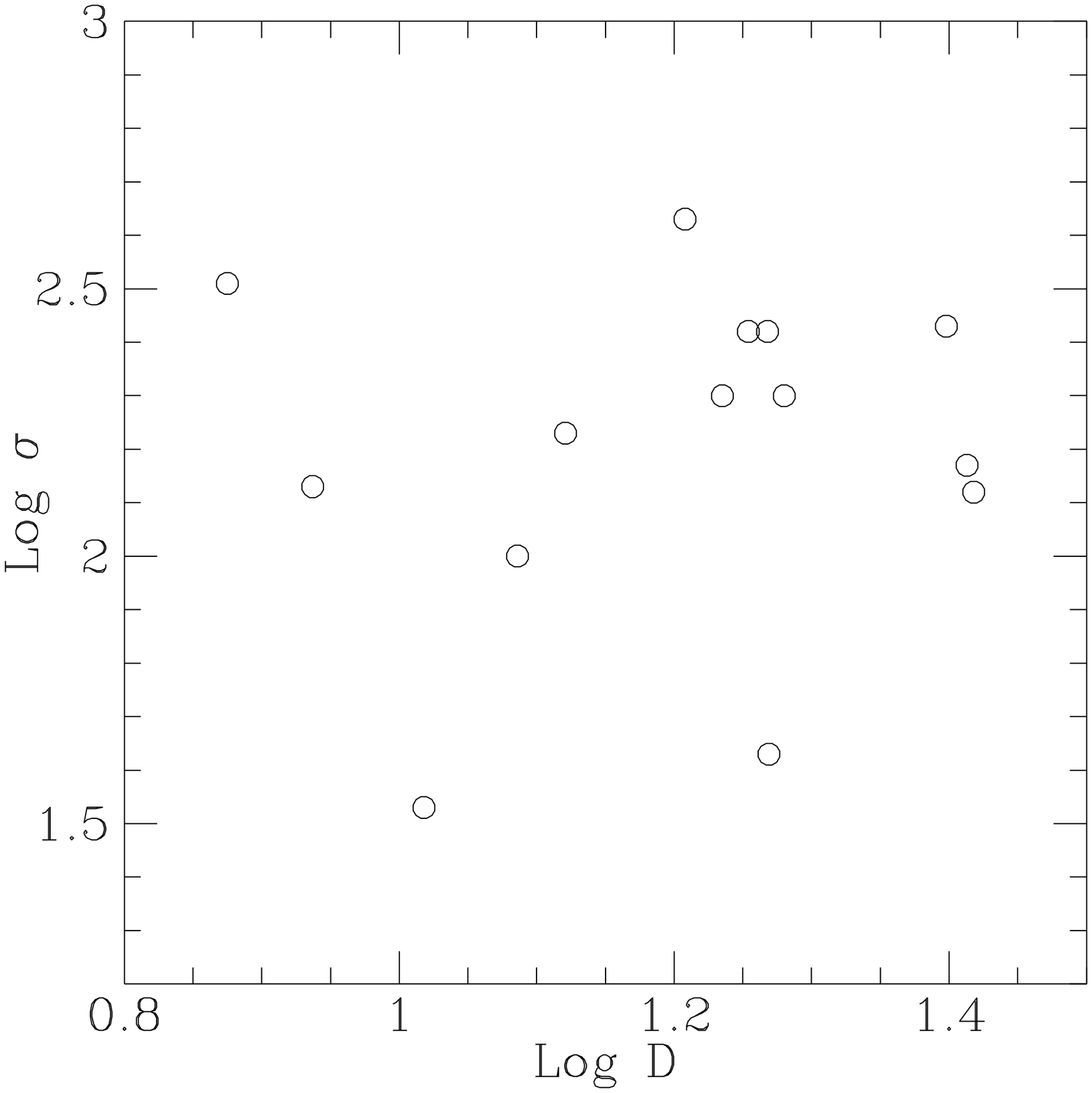,height=8cm,width=8cm}}
\caption{ Log $D$ versus $\sigma $ for
the 14 largest spirals belonging to the groups in our sample which host at
least one elliptical or lenticular.}
\end{center}
\end{minipage} \hfill
\end{figure}

\bigskip

These different behaviors of early and late type objects can have different
explanations. One could argue that spiral dominated groups are more likely \
than early-type ones to be chance alignments in filaments or clusters (e.g.
Hernquist et al. 1995, Pildis 1995). This, however, is not the case since we
do not find any correlation between $Log\;D$\ and $Log\;\sigma $ even if we
restrict our late type sample to the 14 which belong to groups in the LEG\
sample (Fig.3). A second possible explanation is that isophotal
diameters are not as good ''physical'' tracers of the evolutionary history
as the effective radii are for the early type objects but this objection can
be ruled out on the basis that almost all late type galaxies in CGs show
signs of ongoing strong interactions and, therefore, should be truncated by
tidal stripping. A third, more likely explanation comes in a rather
straightforward way from the SGS: late type objects are mainly just now
infalling and are therefore poor tracers of the gravity potential.

\bigskip

One more word of caution:\ when dealing with HCGs, distance dependent
selection effect need always to be checked (Whitmore 1990). However: i) if
we take into account such correlation and detrend our data for the distance
effects we still find a strong correlation (confidence level larger than $%
99.5\%$);\ ii) distance dependent selection effects should work in the same
way on both effective radii and isophotal diameters and we do not find any
correlation for the 14 late type galaxies in the LEG\ sample. \smallskip
Since this may be related to the use of $D$ instead of $R_e$, we checked the
dependence of $Log(D)$-$Log(\sigma )$ for the LEG finding a correlation
similar to that in Eq\ref{Eq. 1}.

\bigskip 
Furthermore, the link between an individual property of the LEG\
(namely the effective radius) which is related to the total luminosity and
hence the mass of the galaxy (Caon et al. 1994)\ and the velocity dispersion
of the groups which is instead a parameter related to the total mass of the
groups, confirms both the physical reality of these groups and the fact that
ellipticals dominate the dynamics of the groups.\\

\vspace{1cm}
{\bf \hspace{2cm}References}\\

\noindent
Bettoni E., Fasano G., 1993, AJ, 105, 1291\\
\noindent
Boselli A., Mendes De Oliveira C., Balkowski C., Cayatte V.,
Casoli F., 1996, A$\&$A, 314, 738\\
\noindent
Capaccioli M., Caon N., D'Onofrio M., 1992, MNRAS, 265, 1013\\
\noindent
Caon N. Capaccioli M., d'Onofrio M., Longo G., 1994, A$\&$A,
268, L39\\
\noindent
Diaferio A., Geller M.J., Ramella M., 1994, AJ, 107, 868\\
\noindent
Ebeling H., Voges W., B$\ddot{o}$hringer H., 1994, ApJ, 436, 44\\
\noindent
Fasano G., Bettoni D., 1994, AJ, 107, 1649\\
\noindent
Governato F., Tozzi P., Cavaliere A., 1996, ApJ, 458, 18\\
\noindent
Hamabe M., Kormendy J., 1987, in IAU Symp. N. 127, ed. T. de
Zeeuw, p. 379\\
\noindent
Hernquist L., Katz N., Weinberg D., 1995, ApJ, 442, 57\\
\noindent
Hickson P., 1982, ApJ, 255, 382\\
\noindent
Hickson P., 1994, Atlas of Compact Groups of Galaxies (Basel:
Gordon and Breach)\\
\noindent
Hickson P., 1997a, in Astrophys. Lett \&\ Commun., in press\\
\noindent
Hickson P., 1997b, ARA\&A, 35, 357\\
\noindent
Hickson P., Kindl E., Auman J.R., 1989, APJSS, 70, 687\\
\noindent
Hickson P., Mendes De Oliveira C., Huchra J.P., and Palumbo
G.G.C., 1992, ApJ, 399, 353\\
\noindent
Hickson P., Richstone, D.O., Turner E.L., 1977, ApJ, 213, 323\\
\noindent
Longo G., et al. in preparation\\
\noindent
Kelm B., Palumbo G.G.C., 1995, Astrophys. Lett. $\&$ Commum.,
31, 299\\
\noindent
Maia M., da Costa N., 1990, ApJ, 350, 457\\
\noindent
Mamon G.A. 1987, ApJ, 321, 622\\
\noindent
Mamon, G.A. 1990, in Paired and Interacting Galaxies,eds.
J.Sulentic and W.C.Keel, Nasa, Washington, P.609\\
\noindent
Mendes de Oliveira C., Hickson P., 1991, ApJ, 380, 30\\
\noindent
Mendes de Oliveira C., Hickson P., 1994, ApJ, 427, 684\\
\noindent
Mulchaey J.S., Zabludoff A.I. 1997, preprint astro-ph 9708139\\
\noindent
Ostriker J.P., Lubin, L.M., Hernquist, L. 1995, ApJ, 444, L61\\
\noindent
P\`{a}ramo J.I., V\'{i}lchez J.M., 1997, ApJ, 489, L13\\
\noindent
Pildis R.A., 1995, Ph. D. Thesis, University of Michigan\\
\noindent
Pildis R.A., Bregman, J.N., and Evrard, A.E. 1995, ApJ, 443, 514\\
\noindent
Ponman T.J., Bourner, P.D.J., Ebeling, H., B$\ddot{o}$hringer H.
1996, MNRAS, 283,690\\
\noindent
Ribeiro A.L.B., de Carvalho R.R., Coziol R., Capelato H.V., Zepf
S.E. 1996, ApJ, 463, L5\\
\noindent
Rubin V.C., Hunter, D.A., Ford, W.K. jr. 1991, ApJS, 76, 153\\
\noindent
Saracco P., Ciliegi, P. 1995, A$\&$A, 301, 348\\
\noindent
Vennik J., Richter G.M., Longo G., 1993, AN, 314, 393\\
\noindent
Whitmore B.C., 1990, Clusters of Galaxies: STSci Symposium no.
4, Oegerle W.R. et al. eds. (Cambridge: Cambridge University Press), P.351\\
\noindent
Whitmore B.C., 1992, in ''Physics of Nearby Galaxies: Nature or
Nurture?'', Thuan T. et al. eds., ed. Fronti\'{e}res\\
\noindent
Williams B.A., van Gorkom J.H., 1988, AJ, 95, 352\\
\noindent
Zabludoff A.I., Mulchaey J.S. 1997, preprint astro-ph 9708132\\
\noindent
Zepf S.E., 1993, ApJ, 407, 448\\
\noindent
Zepf S.E., Whitmore, B.C., Levison, H.F. 1991, ApJ, 383,524\\
\noindent
Zepf S.E., Whitmore, B.C., 1993, ApJ, 418,72\\

%

\address
A.A.Shaker, G.Longo, P.Merluzzi\\
Osservatorio Astronomico di Capodimonte\\
Via Moiariello, 16\\
I -- 80131 Napoli\\
ItalyEND

\bigskip

\noindent
e-mail addresses:\\
shaker@astrna.na.astro.it\\
longo@astrna.na.astro.it\\
merluzzi@astrna.na.astro.it\\

\end{document}